\documentclass[,final]{aipproc}

\layoutstyle{8x11double}

\begin{document}

\title{TeV Gamma-ray Astronomy: The Story So Far}

\classification{97.60.Bw, 97.60.Gb, 97.60.Lf, 98.70.Sa, 98.54.Cm}

\keywords {TeV $\gamma$-Ray Observations}

\author{Trevor C. Weekes}{ address={Fred Lawrence Whipple
    Observatory, Harvard-Smithsonian Center for
Astrophysics, Amado,
    AZ 85645, USA},altaddress={tweekes@cfa.harvard.edu} }

\begin{abstract}
 A snapshot is presented of the present status of our knowledge
of the TeV gamma-ray universe. Emphasis is put on observations
made using the imaging atmospheric Cherenkov technique. The
capabilities of the present generation of telescopes is listed.
Progress has been dramatic and several features have been 
different from what was anticipated. The catalog of sources
includes some 78 objects and these are tabulated as extragalactic 
sources (24), supernovae remnants (11), pulsar wind nebulae (10),
binaries (4), miscellaneous (9), diffuse high energy sources (3) and unidentified 
sources (20). Some comments are made on the factors influencing the
past and future development of the field.
\end{abstract}

\maketitle


\section{Status of TeV Gamma-ray Astronomy, c. 2008}

Having had the honor of presenting status
reports at two of the last three Heidelberg meetings
on TeV gamma-ray astronomy
\cite{weekes93}, \cite{weekes00}, I am painfully aware (with
the benefit of hindsight) of my
shortcomings at such a task, in particular at my attempts
to predict the future growth of the field where I have
generally erred on the conservative side and have been
pleasantly surprised by the pace of discovery.

Hence here I will confine myself to reporting on the status of
the field as I know it today (July, 2008) which will
therefore rely mostly on my knowledge of the published
papers and recent excellent reviews \cite{hinton07}, 
\cite{deangelis08}, \cite{aharonian08} as well as the
comprehensive White Paper on TeV ground-based gamma-ray astronomy
that was recently put together by members of the US TeV gamma-ray
community \cite{buckley08}. I will attempt
to summarize the status and capability of the various
observatories, outline the sensitivity that can be achieved
with existing instrumentation, the range of observed
phenomena, make some attempt to catalog the credible
discrete sources reported to date, and provide some personal
perspective on the progress of, and prospects for, the field. 
I will make no attempt
to describe the significant improvements that are planned for
existing observatories, some of which are in an advanced
state of construction.

\subsection*{Caveat}

Nothing dates more rapidly than a written status report.
This is particularly true of one presented at the beginning
of a symposium where the most exciting results from the
various groups have been embargoed so that they can be
presented at the symposium. 

Thus even before the ink on the report is dry, it is out-of-date. 
Its value therefore is merely to serve as a historical
benchmark and to summarize what one of us thought he knew at the
start of the symposium and to provide some kind of reference
point for future developments.

\section*{What we have today}

\subsection*{The Atmospheric Cherenkov Technique}

With the notable exception of the remarkable results from
the Milagro experiment \cite{abdo07}, the bulk of the
observational results at TeV energies have come from
telescopes using the atmospheric Cherenkov technique.
Although the basic technique was developed some fifty years
ago, it was not until the development of the so-called
imaging atmospheric Cherenkov technique (IACT) that the
first indication of a credible detection was apparent
\cite{cawley85}, \cite{weekes89}. Given the rather murky history that
has characterized the early results at all gamma-ray
energies (100 MeV energies as well as 1 TeV energies), it is
not surprising that this early detection was treated with
some skepticism. 

In its earliest manifestation the technique
was characterized by its:

\begin{itemize}
\item Simplicity
\item Economy
\item Elegance
\end{itemize}

\begin{figure}
\includegraphics[height=.3\textheight]{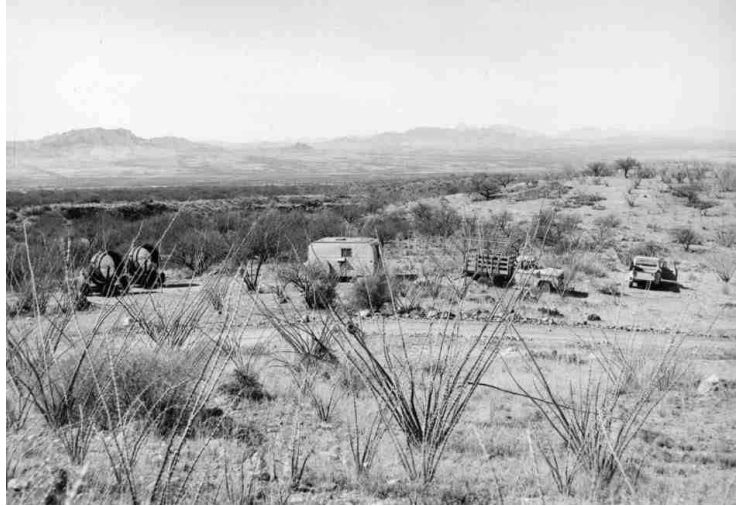}
\caption{The first TeV gamma-ray observatory in the United States
consisted of two 1.5m telescopes (made from World War II 
searchlight reflectors) above (left center); 
the telescopes were manually operated and were 
located at a dark site in southern Arizona during the winter of 
1967-8 \cite{fazio68}. The telescopes were directed (by eye) at
a point ahead of the position of the putative source so that the
earth's rotation swept the source through the field of view. 
Power came from an electric generator on
the back of the truck (center right) and 
the pulse counting electronics were housed
in a small trailer (center). The system was mercifully free of
computers and the analysis was done offline with a mechanical
calculator. No sources were detected. }
\label{searchlights}
\end{figure}

Although the early experiments (an example, the first Whipple
Observatory experiment shown in Figure \ref{searchlights})
did not succeed in producing convincing evidence for the
existence of any sources, they showed that it was possible for
small groups to participate in the exciting new field of high
energy astrophysics; this field was soon to be dominated by
elaborate and expensive experiments in balloons 
or on satellites. The elegance of
the Cherenkov technique was apparent in its economy as to the energy
intercepted that was necessary to detect the gamma ray; as
Ken Greisen pointed out, the ground-based technique is remarkable in
that only one millionth of the energy of the primary gamma
ray (in the form of Cherenkov light photons) need be
collected by the telescope for the gamma ray to be detected.

\begin{figure}
\includegraphics[height=.2\textheight]{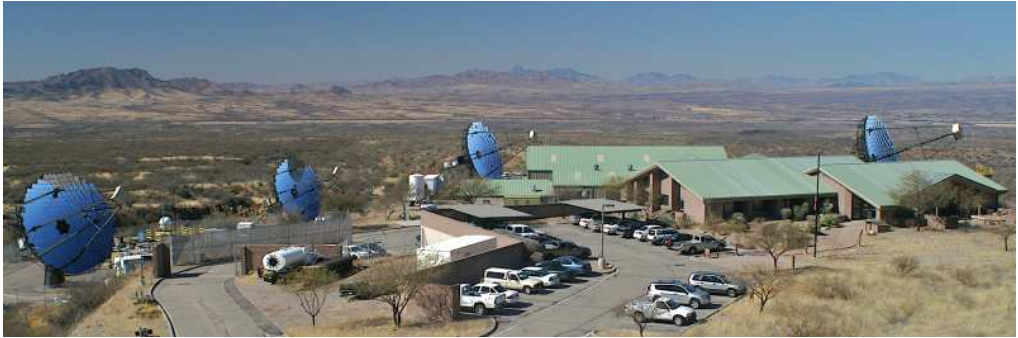}
\caption{The VERITAS observatory, the newest of the third generation 
IACT observatories,
saw first light in April, 2007. Note that VERITAS is in the exact
same location as the telescopes shown in Figure \ref{searchlights}.
Each of the four telescopes has an aperture of 12m (collection area of
106m$^{2}$) and a camera with 499 pixels.}
\label{VERITAS}
\end{figure}

The imaging atmospheric Cherenkov technique as practiced
today (Figure \ref{VERITAS}) with its multitude of pixels,
multiple large optical reflectors, and high speed data
acquisition systems is certainly not simple. Of necessity,
the costs of such systems are now large and prohibitive for
small research programs. The typical state-of-the-art observatory
costs \$20M and authorship lists are in the hundreds. 
Only the elegance survives and
still makes the technique attractive to cosmic ray
physicists and refugees from large high energy physics
experiments.

With the current emphasis on "green" technology it is
perhaps worth noting that the detection of high energy
particles using their secondary emissions in the atmosphere
is inherently a "green" technique since in no way is the
natural path of the gamma ray, destined to cross the
wilderness of interstellar and perhaps intergalactic space 
and end its life by collision with an
air molecule, disturbed. The gamma ray is not even aware it
has been detected! In contrast the highly technical 
gamma-ray space telescope must intercept, and destroy, the primary
gamma ray in its complex silicon layers; it thus ends its
life prematurely and catastrophically and never completes its
original destiny.


\subsection*{The Tools Available Today}

Although all ground-based experiments that have the sensitivity
to detect a source like the Crab Nebula can contribute to TeV
gamma-ray astronomy, the recent literature (and this symposium)
 has tended to be 
dominated by the "Big Three", H.E.S.S., MAGIC and
VERITAS. However CANGAROO III, the Whipple 10m telescope
 and the recently completed HAGAR in Ladhak, India also play an
important role, particularly in monitoring variable sources.
Some of the characteristics of these observatories are listed in Table
\ref{observatories}.

It should be noted that only three of the original four 
telescopes of CANGAROO III are now in operation so the sensitivity
is reduced from the original designed threshold (M. Mori,
private communication). The threshold shown for H.E.S.S. is that 
at first light and when most of their pioneering discoveries were
made; because of mirror weathering the threshold
is now higher. HAGAR has only recently come
on-line so its sensitivity has still to be confirmed; this value
and those for the other two Indian experiments were supplied by
B.S. Acharya (private communication). HAGAR is noteworthy in that 
it will be the first telescope to operate at an elevation $>$ 4km.
Of the eight observatories listed, all but HAGAR and PACT use the
IACT; their approach is to use wavefront sampling with an array
of small telescopes.  
\begin{table}
\begin{tabular}{lrrrrrrr}
\hline
\tablehead{1}{r}{b}{Observatory}
& \tablehead{1}{r}{b}{Elevation\\(km)}
& \tablehead{1}{r}{b}{Telescopes\\ \#}
& \tablehead{1}{r}{b}{Mirror Area\\($m^2$)}
& \tablehead{1}{r}{b}{FoV\\(degrees)}
& \tablehead{1}{r}{b}{First Light}
& \tablehead{1}{r}{b}{Threshold\\(GeV)}
& \tablehead{1}{r}{b}{Sensitivity\\(\%Crab)}\\
\hline
H.E.S.S. & 1.8 & 4 & 428 & 5 & 2003 & 100 & 0.7\\
VERITAS & 1.3 & 4 & 424 & 3.5 & 2007 &100 & 1\\
MAGIC & 2.2 & 1 & 236 & 3.5 & 2005 & 50 & 1.6 \\
HAGAR & 4.3 & 7 & 31 & 3  & 2008 & 60 & 9\\
Whipple & 2.3 & 1 & 75 & 2.2  & 1985 & 400 & 10 \\
CANGAROO III & 0.1 & 3(4) & 172  (230) & 4 &2006 & 400 & 10\\
PACT & 1.1  & 24 & 107 & 3 & 2001 &  750 & 11\\
TACTIC & 1.3 & 1 & 10 & 2.8 & 2001 & 1500 & 70 \\
SHALON & 3.3 & 1 & 11.2 & 8 & 1996 & 1000? & ? \\	 
\hline
\end{tabular}
\caption{Major Existing ACT Facilities}
\label{observatories}
\end{table}

\section*{The Present Capabilities}

\subsection*{Wide Spectral Energy Range : 25 GeV to 100 TeV}

\begin{figure}
  \includegraphics[height=.4\textheight]{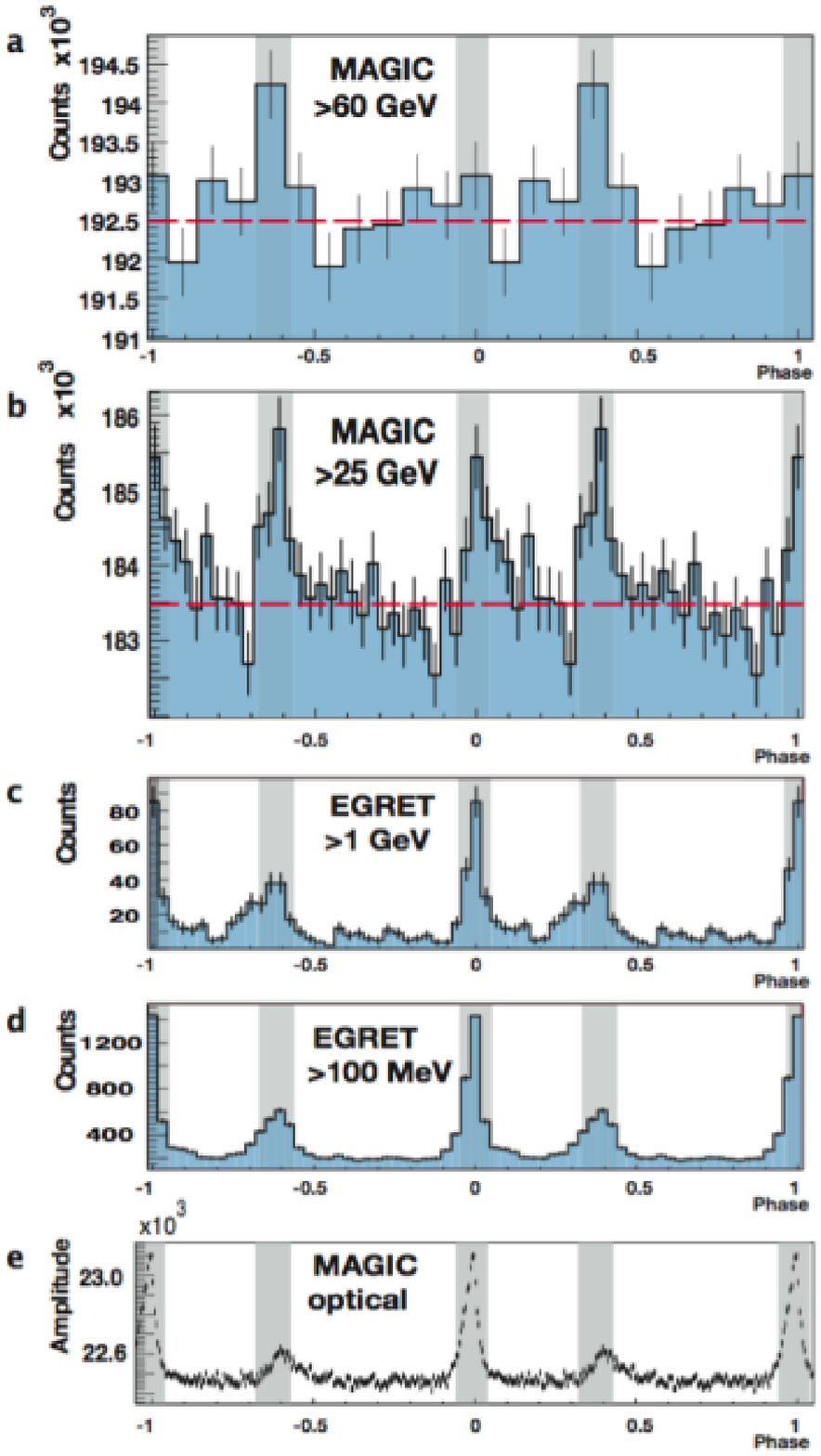}
  \caption{
The first indication of a weak pulsed signal from the Crab pulsar
at energies above 60 GeV from the MAGIC group \cite{otte07}.
A later report (this symposium) confirmed the signal and extended
the observations down to 25 GeV, the lowest energy at which a
signal has been reported using the atmospheric Cherenkov technique.
}
  \label{pulsar}
\end{figure}

\begin{figure}
\includegraphics[height=.28\textheight]{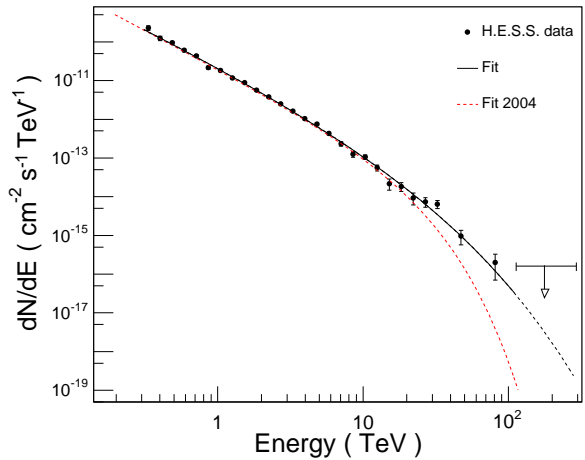}
 \caption{
The H.E.S.S. gamma-ray spectrum of RX~J1713.7-3946, the strongest
gamma-ray source in the Southern Hemisphere \cite{hessrwj}.
The data points can be fit by a power law
      with exponential cutoff.  
     The upper limit, indicated by the black arrow,
      covers the energy range from 113 to 300~TeV.
Particle acceleration up to at least 100~TeV is inferred from these
observations; although the progenitor particles could be hadrons,
electron progenitors are not ruled out.
}
  \label{highE}
\end{figure}

\begin{figure}
\includegraphics[height=.28\textheight]{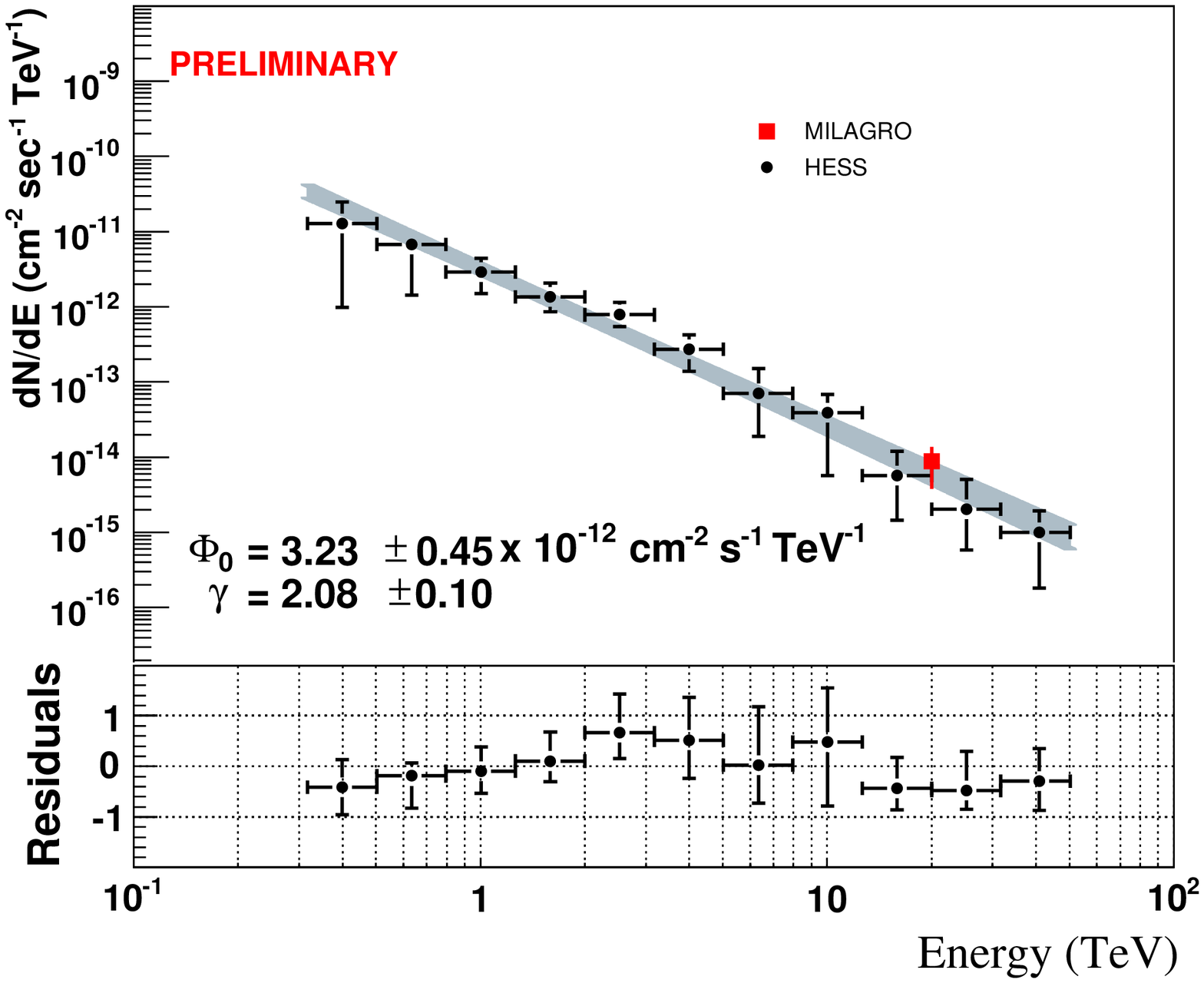}
 \caption{
The differential energy spectrum of the extended source which
was first reported by Milagro \cite{abdo07}. The spectrum shown
here is as detected by H.E.S.S. \cite{hess1908}
which saw it as a hard spectrum source above 300 GeV. The H.E.S.S.
source is extended, with a FWHM size of
0.5$^{\circ}$.
The single differential point seen by Milagro at 20 TeV is shown as a square.
}
 \label{Milagromap}
\end{figure}

The earliest experiments had energy thresholds in excess of
5 TeV. As the size of  the telescopes has increased and the
sophistication of the triggering improved, the energy
threshold has steadily dropped, so that now results are
presented with thresholds as low as 25 GeV (Figure
\ref{pulsar}). The upper energy bound is determined by
exposure time and can be extended by observing at low
elevations where the collection area and energy threshold
increase. It is certainly possible to make observations with
telescopes using the IACT up to energies of 100 TeV (Figure \ref{highE},
Figure \ref{Milagromap}).
The motivation to go to lower energies comes mainly from the 
desire to study distant AGN, which are expected to have soft spectra,
and Gamma Ray Bursts and pulsars for the same reason. Higher energies
are particularly important in the study of Supernovae Remnants
since such observations have the best hope of separating out
the contributions from hadronic and electronic progenitors. 
Although most IACT observatories strive to achieve the lowest
possible energy thresholds, in practice the technique is still
most sensitive at energies around 200 GeV and this is where
most of the new sources have been discovered. The detection of
the Crab pulsar (this symposium) is an obvious exception. 

\subsection*{Energy resolution: 10 to 35\%}

Once a source has been detected, interest centers on the determination
of the energy spectrum. Most sources can be fit with a simple
power law. Structure in the spectrum can be an important clue to 
the emission mechanism and possible absorption processes. Energy
resolution can range from as little as 10\% with arrays of telescopes, 
which permit the determination of the impact parameter, to
30-40\% with single telescopes. The signal strength should be 
at least 5$\sigma$ for a meaningful measurement. At low energies
spectral mesurements are limited by threshold effects and at high
energies by statistics. A steep spectrum must always be treated with
caution since this is the characteristic of a false detection which
is usually caused by an uneven sky background effect.

Energy resolution is important to those who see TeV astronomy
as a window in which to explore dark matter; a source with
high density might be expected to have a high concentration of
neutralinos and might be identified by a line in the 100-1000
GeV range. This possibility is a driving force for many TeV
scientists with a background in high energy physics.

\subsection*{Flux Range: 1 to 1500\% of the Crab}

\begin{figure}
\includegraphics[height=.3\textheight]{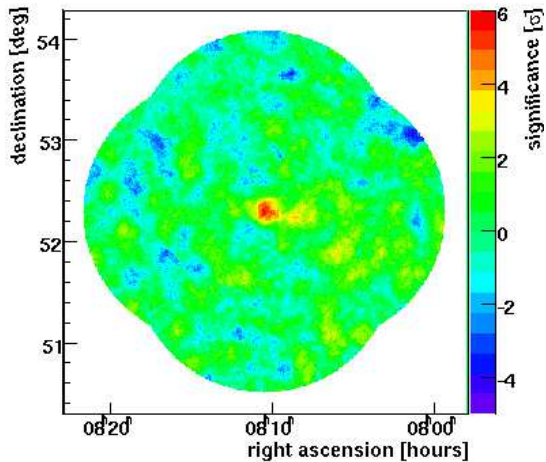}
 \caption{
VERITAS observations of the HBL blazar 1ES0806+514 (z=0.138) show 
weak but steady emission above 300 GeV \cite{atel0806}; 
\cite{cogan08}.
Observations were carried out during the construction
of VERITAS and incorporate data using two, three and
four telescopes. This is an example of a detection at
approximately 1\% of the Crab Nebula flux.
}
 \label{weaksource}
\end{figure}

\begin{figure}
\includegraphics[height=.17\textheight]{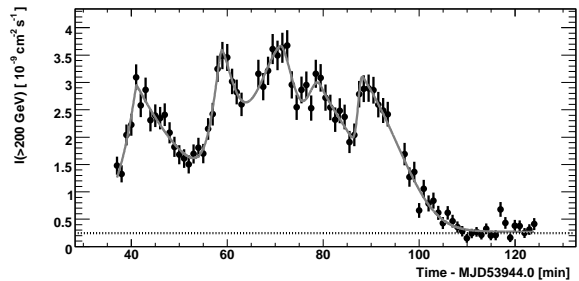}
 \caption{
The extraordinary outburst from PKS2155-302 observed by H.E.S.S.
in 2006 \cite{bigflare}. At peak the flux was 15 times 
that of the Crab Nebula.  
Short-term variability has also been seen from other AGN: Markarian 421
by Whipple and VERITAS and Markarian 501 by MAGIC.
The time-scales of these bursts are among the fastest ever
seen in blazars at any wavelength.  
}
 \label{bigflare}
\end{figure}

The Crab Nebula is the strongest steady TeV source in the
sky; with a declination of +22$^{o}$ it is visible from
both hemispheres and has a moderately hard spectrum. Thus it
is useful as a standard candle for comparing instrument
sensitivities and source strengths. With integration times
of 50 hours (a sizable fraction of the observing year for
most observatories), sources can be detected that are 1\% of
the Crab (Figure \ref{weaksource}). However most of the
reported sources have signal strengths well in excess of
1\%. Systematics tend to limit longer integration times and
hence the detection of weaker sources. Flaring AGN have been
detected with fluxes (for short periods) in excess of
fifteen times the Crab (Figure \ref{bigflare}).

The usual standard for acceptance of a new source is that the
signal should be at the 5$\sigma$ level. This is a fairly
conservative criteria and is probably justified when the
systematics are not fully understood.  However observations
of important candidate objects which are above the 4$\sigma$ level
should certainly be reported but treated with some caution.
Independent verification by another observatory is perhaps 
the best criterion for credibility but if that standard was
adopted the TeV source list would be very short. This is
particularly so in the Southern Hemisphere where most of the
sources have been seen by just one experiment.

\subsection*{Angular Resolution: 2 arc-min to 3 degrees}

\begin{figure}
 \includegraphics[height=.3\textheight]{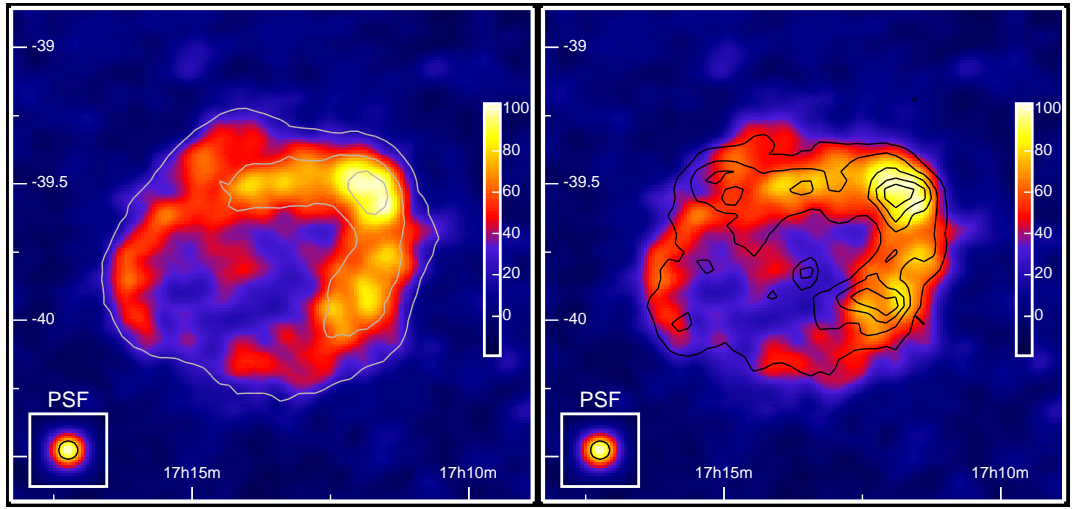}
  \caption{
 These are two remarkable representations of the H.E.S.S. observations of RX~J1713.7-3946
 \cite{hessrwj}.
    The images are smoothed with a Gaussian
      of 2 arc-min.  On the left, the overlaid
      light contours show the significance levels of the different
      features. The levels are at 8, 18, and 24$\sigma$. On the
      right the X-ray ASCA contours are shown as black
      lines.  The full detail can only be
seen in the color version of the figure in the original publication}
 \label{RJW1713map}
\end{figure}

The IACT is optimized for point source detection. H.E.S.S. has an
angular resolution of 2 arc-min as demonstrated in the
beautiful map of RX~J1713.7-3946 (Figure
\ref{RJW1713map}). In general, extended sources are
more difficult to detect by IACT telescopes. In the surveys
that have been made of the Galactic Plane it is clear that
many Galactic sources are not point-like. The Milagro
experiment has poorer resolution but greater sensitivity for
the detection of extended sources (Figure \ref{Milagromap}).
The source location capability is usually sufficiently
good that there is no ambiguity in the identification with
the target object. Unlike the 100 MeV region where the gamma-ray
point source sensitivity is severely limited by the contribution
from the Galactic Plane, TeV observations have basically the same
sensitivity over all the sky and hence offer better 
opportunities for source identification for sources that
are detected in both energy bands.

\begin{figure}
\includegraphics[height=.35\textheight,angle=270]{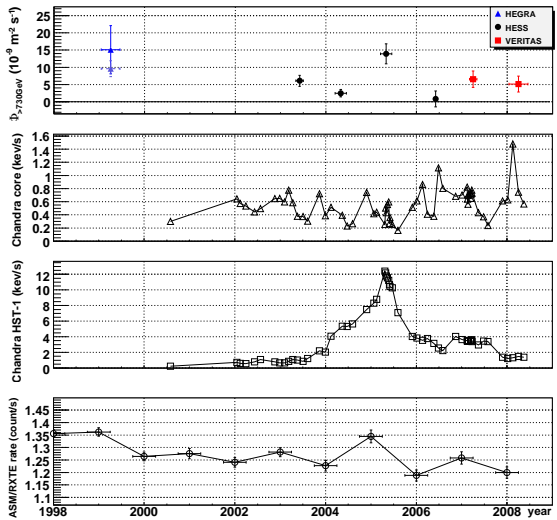}
  \caption{
Top panel: the ten year light curve of
M87 in TeV gamma rays and soft X-rays \cite{hui08}. The TeV 
gamma-ray points are from
HEGRA, H.E.S.S. and VERITAS. Second and third panels: 
X-ray data from Chandra 
(D. E. Harris, private communication) from the core and knot, HST-1.
Bottom panel: X-ray data from the ASM/RXTE quick-look web page. Although
this gamma-ray light-curve only shows the variations on a one year 
time-scale, the gamma-ray
flux has also been seen to vary on a time scale of days.}
  \label{M87lightcurve}
\end{figure}

\begin{figure}
  \includegraphics[height=.16\textheight]{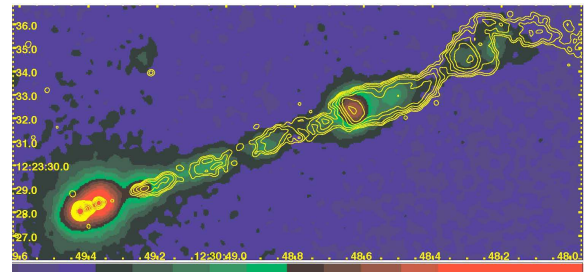}
  \caption{The X-ray image of the jet in M87 as seen by Chandra
(D. Harris, private communication). Radio
contours are superimposed. Interest has centered on the core of the
object (bottom left hand corner) which is resolved in X-rays into
two objects, the core and the close knot called HST-1. Initially
it was thought that the TeV variations were correlated with the X-ray
emission from HST-1 (in 2005) but more recently it appears that it is
correlated with the core. }
  \label{M87image}
\end{figure}

In some cases correlated time variability
at other wavelengths with superior angular resolution can
lead to source identification on the sub-arc-min scale,
e.g. in M87 (Figure \ref{M87lightcurve}, Figure \ref{M87image}).

\subsection*{Distance: 500 ly to 1 billion ly}

\begin{figure}
\includegraphics[height=.3\textheight]{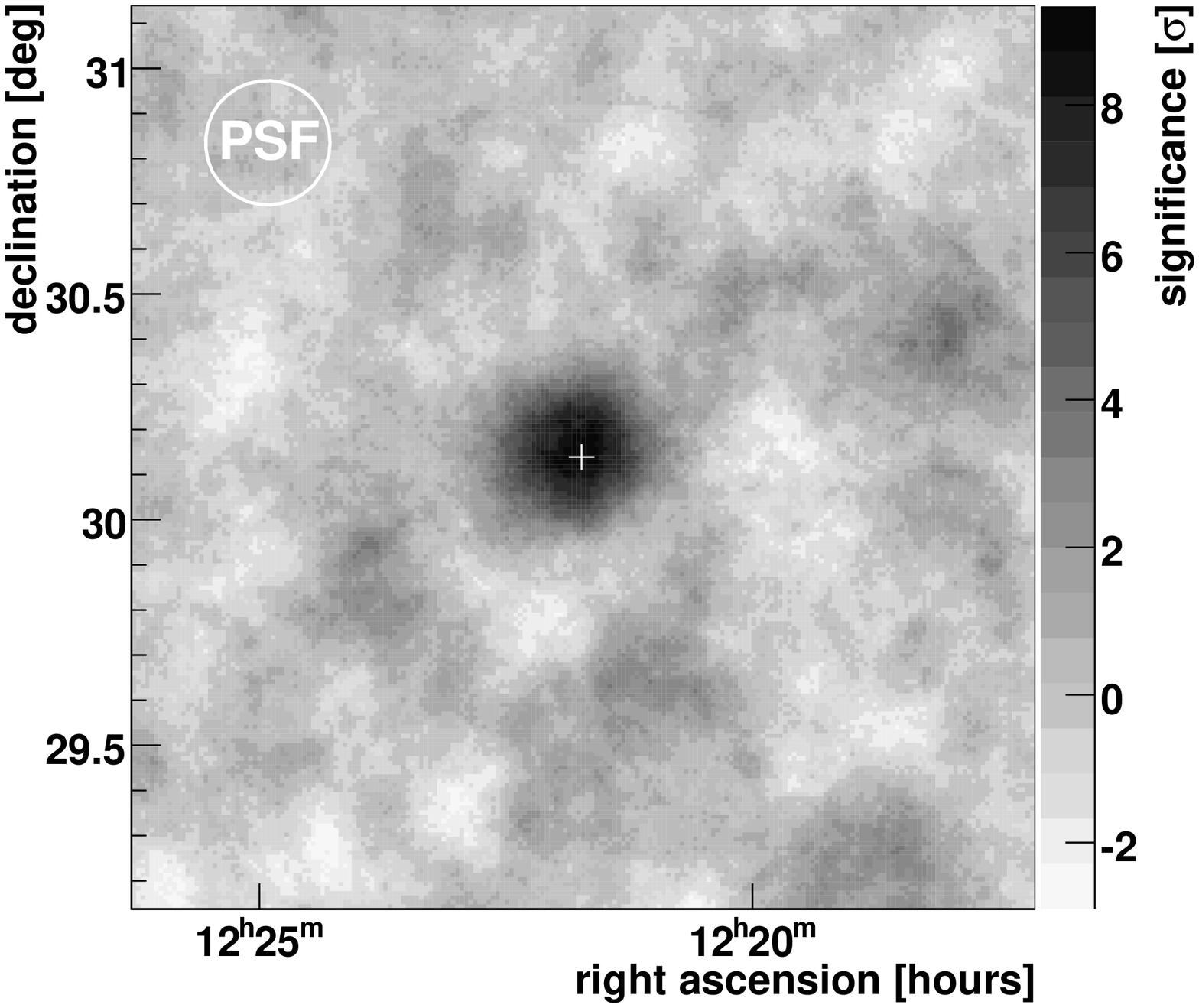}
 \caption{A map of the
significances around the region of the distant HBL AGN, 1ES
1218+304 (z=0.182) as seen by VERITAS \cite{fortin08}. 
The white cross indicates the position of the radio
source. The white circle in the upper left corner shows the angular
resolution of VERITAS (PSF). 
}
  \label{1ES1218map}
\end{figure}

\begin{figure}
\includegraphics[height=.32\textheight]{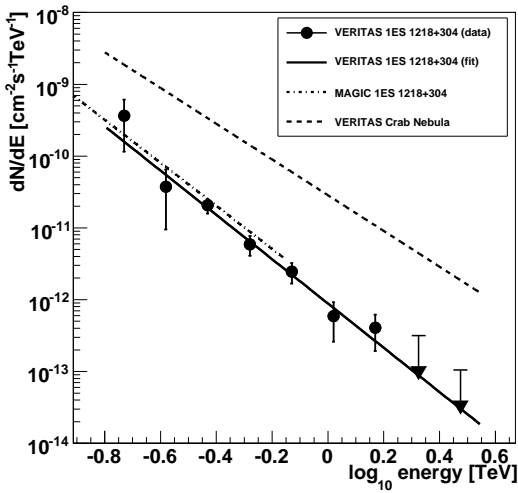}
 \caption{
The differential energy spectrum of 1ES1218+304 measured by VERITAS
with a power law fit shown by the continuous straight line
\cite{fortin08}. The dot-dash
line shows the spectrum measured by MAGIC in their discovery observations
\cite{magic1218}. The dash line is the spectrum of the Crab Nebula 
as measured by VERITAS.
}
  \label{1ES1218spectrum}
\end{figure}

The closest source reported may be Geminga which was nominated
as a candidate source by the Milagro group \cite{abdo07}. 
Blazars are the
most distant objects (Figure \ref{1ES1218map}) that have
been detected. The most distant source detected is still somewhat
controversial since there is always uncertaintly about the
reshifts of BL Lac objects because of their paucity of emission
lines.  One would expect the observed
spectra of AGN to soften with increasing redshift since the infrared
absorption effect should increase with energy; this is generally
the case (Figure \ref{1ES1218spectrum}). Certainly there are
several AGN detected with redshifts in excess of 0.2; the most
distant object is probably 3C279 but this detection still
awaits confirmation. 3C66a, which was originally detected by the
Crimean Astrophysical Observatory group \cite{neshpor98} and 
has recently been confirmed by VERITAS \cite{atel3C66a}, is reported 
to have a redshift of 0.444.

Because the IACT has very good flux sensitivity for the detection
of short transients, e.g. Gamma Ray Bursts, it may be that these
will be the most distant detectable sources of TeV gamma rays.

\subsection*{Time Variations: minutes to years}

\begin{figure}
\includegraphics[height=.4\textheight]{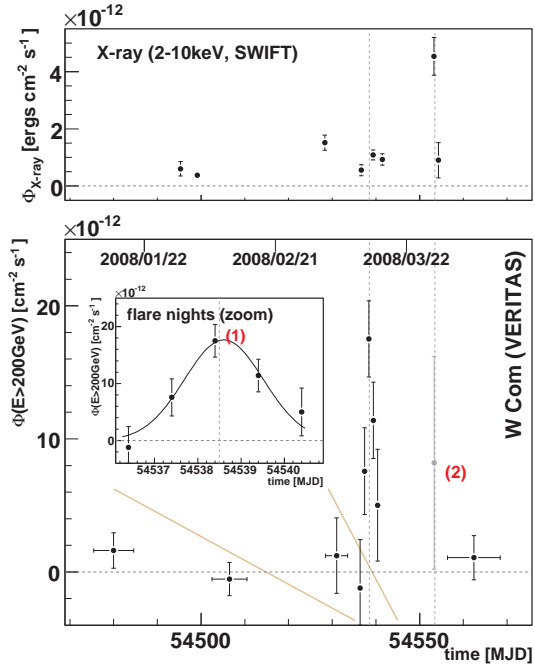}
\caption{
The upper panel shows the
light curve of the integral photon flux from 
W Comae as observed by VERITAS in March, 2008 \cite{atelwcomae1},
\cite{beilicke08}.
Each point corresponds to one night of observation. 
The upper panel shows the X-ray flux as measured by Swift. This
was the first detection of a signal from this IBL AGN. A second
stronger flare was seen in June, 2008 \cite{atelwcomae2}.
The source is only detected when flaring.}
\label{wcomae}
\end{figure}

Although some TeV sources have been observed to exhibit
rapid time variability, the majority of the reported TeV sources
exhibit steady emission within the sensitivity limits of the
observatories and the duration of the observations. Time
variations have been observed on a scale as short as 2-3
minutes (Figure \ref{bigflare}) and as long as years (Figure
\ref{M87lightcurve}). Some sources are only detected when a
flare occurs, e.g. W Comae (Figure \ref{wcomae}). The variability
of the AGN sources makes TeV astronomy more interesting; it is
impossible to predict when a source like Markarian 421 will
be flaring so that on-line data analysis is always exciting
and often enough to keep the observer awake during a long
night of observing. In a few cases optical brightening of the
AGN has triggered the detection of TeV flaring AGN. 
The Galactic sources are
generally steady and predictable; 
the exceptions are the periodic pulsar and binary
sources. 

The time-scale of the variations of the AGN signals
limits the size of
the emitting region and can be an important probe for
cosmological and fundamental physics studies.
Doppler factors in excess of 100 are required to explain
the observed variability if one assumes the emission
region has a size comparable to the Schwarzschild radius
of a massive black hole.

\subsection*{Multiwavelength Coverage: $10^8 to 10^{27}Hz$}

\begin{figure}
\includegraphics[height=.23\textheight]{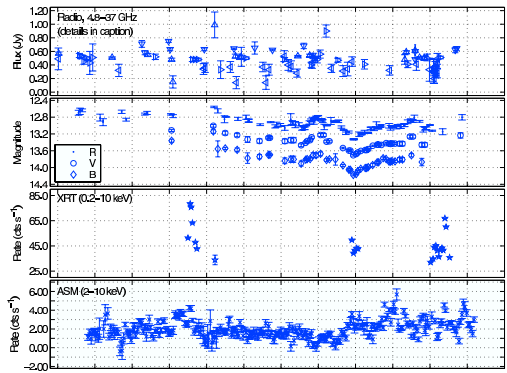}
  \caption{Observations of Markarian 421 taken in an extensive
multi-wavelength campaign from November, 2005 to June, 2006,
which was organized by the Whipple gamma-ray group.
Data were taken on four radio telescopes and thirteen optical 
telescopes (as well as two X-ray satellites and the Whipple 10m
telescope) (Figure \ref{horanmrk421b}). 
Variability was found at almost all wavelengths.
 The radio data
are plotted in the top panel: 4.8 GHz, 8 GHz, 14.5 GHz and 
37 GHz. The
optical data are combined from the different observatories in three color
bands.
}
 \label{horanmrk421a}
\end{figure}

\begin{figure}
\includegraphics[height=.19\textheight]{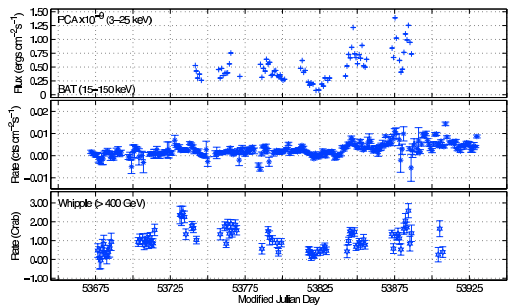}
\caption{
The X-ray and TeV gamma-ray data points over the same time
period as in Figure \ref{horanmrk421a}  . 
The X-ray points come from
RXTE (ASM) and Swift (BAT). The gamma-ray points are from the
Whipple 10m gamma-ray telescope with threshold 400 GeV; each data
point is the flux recorded on that night of the observing campaign.
}
  \label{horanmrk421b}
\end{figure}

To probe the astrophysics of the sources it is particularly
valuable to make correlated observations across the
electromagnetic spectrum. The TeV observations, coming at
the extreme end of the spectrum, are unique and generally
stretch the models to their limits. A number of campaigns
have been organized around the TeV observations; these
typically involve radio, infrared, optical, 
X-ray and gamma-ray observatories; an example of the results
of one such campaign (on the most variable AGN, 
Markarian 421) is shown in Figure
\ref{horanmrk421a} and \ref{horanmrk421b} (D. Horan, private
communication). Of particular 
value in the study of
blazers are correlated observations with hard X-ray
observatories.

The organization of multi-wavelength campaigns is particularly
difficult because of the difficulties of scheduling diverse
instruments in space and on the ground. It is complicated
by the different cultures that prevail in the different
wavebands. The production of papers based on such campaigns
requires the patience of Job. Since many TeV physicists do 
not have a background in classical astronomy, involvement 
in such campaigns has the added advantage of increased awareness
in other astronomical disciplines. It also increases awareness of
TeV astronomy in the wider astronomical community.

\subsection*{Source Density}

\begin{figure}.
\includegraphics[height=.32\textheight]{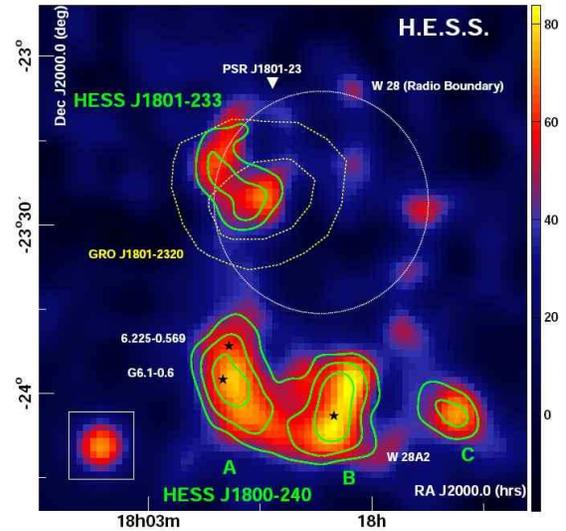}
 \caption{The TeV gamma-ray complex image near W28 
as seen by H.E.S.S. \cite{hessw28}. 
The solid contours of the gamma-ray excess are at  
    significance levels of 4, 5, and 6$\sigma$. 
The dashed circle depicts the approximate radio boundary of the SNR W~28.
There are at least four gamma-ray sources in this map.}
 \label{W28map}
\end{figure}

\begin{figure}
\includegraphics[height=.32\textheight]{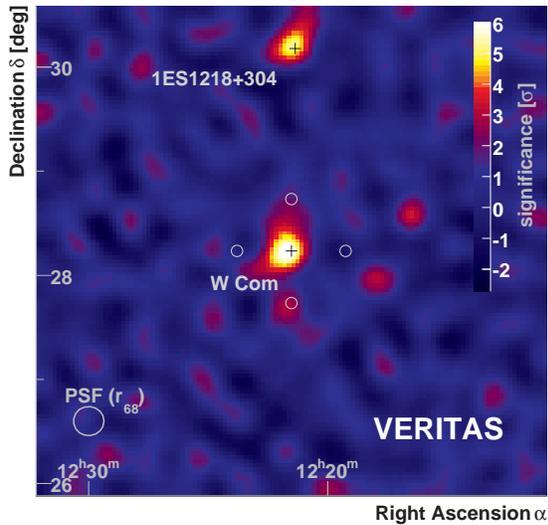}
\caption{
VERITAS sky map of significances centered on the position of
W Comae (cross) \cite{beilicke08}. Although 1ES1218+304 
(z=0.182) had previously
been detected, it was again apparent in the W Comae (z=0.0.080)
observations
and illustrates the sensitivity to off-axis sources and 
the density in the sky of AGN.}
\label{wcomae1es1218}
\end{figure}

Source confusion was not considered a problem for TeV
observatories until recently when the unexpected density 
and complexity (Figure \ref{W28map}) \cite{hessw28} of
sources indicated that as the sensitivity of observatories
increased, the probability of finding more than one source
within the limited field of view of the detectors was
definitely finite. While this was not too unexpected close
to the Galactic Plane it was somewhat
unexpected when it involved extragalactic sources (Figure
\ref{wcomae1es1218}).

Much of the sky has not yet been systematically surveyed; it
is important to reexamine archival data in cases where
a new source is detected in an apparently empty field.
 
\section*{Progress to date}

Those who have been involved in, or who have followed,
 the development of this field cannot help
but be pleased at the progress made to date. Some of this
progress was predictable but generally I believe most
 observers have been amazed (and pleased)
at how much has been achieved.   
 
A brief list of the highpoints of what we
now know about the TeV sky must include:

\begin{itemize}

\item Multitude of Sources: initially it was supposed that
the TeV sky would be dominated by a small number of
sources, probably supernovae remnants, in which hadrons
would be accelerated by some process and gamma rays would be
produced in the decay of neutral pions created in the
collision of the hadrons with gas; this is clearly not the
case with many diverse categories of sources and with
electron progenitors apparently playing a major role.
\item Supernovae Remnants: as expected, gamma-ray emission
is seen from some supernovae remnants of the shell variety  
but they are by no
means the dominant sources in the Galaxy; the conventional
theory of origin of the cosmic radiation is not
substantiated by the TeV gamma-ray observations made to
date.
\item The 100 MeV Connection: an early assumption was that
the TeV sky would be a weak extension of the rich 100 MeV
sky revealed by SAS-2, COS-B, and EGRET; in fact the TeV sky
has been shown to be an entirely different animal and the
large EGRET catalog is seen to be a poor predictor of TeV
emission. Hard X-ray emission correlates better
with TeV emission and is an indication that Compton-
Synchrotron models with electron progenitors can explain the
emission in many sources.
\item Unidentified Sources: the existence of the so-called
"dark sources" highlights the fact that the TeV sky has
unique features that are not apparently duplicated in other
wavebands. Based on their spatial distribution it appears
that they are mostly Galactic sources.
\item Spectra: the spectra of Galactic TeV sources is, in
general, much harder than expected; with a few notable
exceptions the most interesting discoveries have come, not at
the lowest threshold energies, but at the medium energies
where the telescopes have the maximum flux sensitivity.
\item Extended Sources: since the IACT is particularly sensitive
for the detection of point sources, it was not surprising
that the first sources detected were all 
point-like. That many of the new Galactic sources have significant
angular extent has been a pleasant surprise and has opened
new avenues of astrophysical investigation with comparison
of source maps at different wavelengths. The angular
resolution of the IACT is surprisingly good and exceeds that
found to date at other gamma-ray energies.
\item Extragalactic Sources: the number of extragalactic
sources is larger than expected; clearly cosmic ray
acceleration is not just a Galactic phenomenon. The TeV
blazer catalog is the most complete catalog of any category
of source and can be expected to grow since even with
current sensitivity only a small fraction of the sky has
been explored. In contrast much of the nearby rich Galactic
sky has been scanned with high sensitivity.
\item Transients: one of the great strengths of the ground-
based techniques is their large collection area and hence
high sensitivity to transient emission. The extragalactic
sources exhibit much more variability than the sources found
thus far in the Galactic Plane. The time-scale of variations
in these sources is much shorter than expected and opens
the possibility of testing physical laws as well as pointing
to some extraordinary astrophysical processes.
\item Absorption: there is less absorption (by
photon-photon pair production) in intergalactic space than
had been expected; the observable gamma-ray universe is
therefore larger than the pessimistic early predictions
would have suggested.
\item Moonlight: the advent of arrays of imaging detectors
with stable trigger systems
permit the observation of gamma-ray sources under moonlight
with only marginally reduced sensitivity.
Although generally the time around full moon is avoided, 
the traditional number of useful observing hours can be 
increased by 30-40\%.
\end{itemize}

\section*{Source Catalogs}

It is not possible to summarize all the observational results
in a single publication although a good attempt can be found
in \cite{aharonian08}. In this field catalogs are difficult
because sources are detected by a variety of instruments and new
sources are constantly being added.  
Unlike space missions, ground-based experiments do not have sharp
turn-on and turn-off dates so that it is hard to say when a catalog
is complete. A useful catalog of the many sources
found by the H.E.S.S. observatory can be found on their Webpage
\cite{hessweb}
but naturally this does not include sources detected by 
other groups and hence is biased
towards the Southern Hemisphere. Another useful web catalog is that
maintained by Deirdre Horan and Scott Wakely \cite{horanwakely} which
only lists those sources that have been accepted for publication.
To give some measure of the progress in the field and to provide
a bench mark for checking future progress, the generally 
accepted sources and their vital parameters are gathered here in a 
series of Tables that are loosely based on the above catalogs.

These are divided into five tables: 
Extragalactic sources (Table \ref{extragalactic}),
Supernovae Remnants (Table \ref{snr}),
Pulsar Wind Nebulae (Table \ref{pwn}),
Milagro Sources (Table \ref{milagro}),
Binaries/Miscellaneous (Table \ref{oddsandends}), and
Dark Sources (Table \ref{dark}). All the sources are designated
by their Right Ascension (2000) and Declination (2000) coordinates
which have been rounded; a TeV prefix is attached to each source 
irrespective of its discoverer. 
  
Table \ref{extragalactic} lists
the catalog of known extragalactic sources;
 not only is it
one of the largest source categories but it is also one of
the least ambiguous and the most homogeneous. The 24 sources
are listed in terms of their catalog and common name, 
their redshift (which determines their place in the table), their
flux at 1 TeV (usually variable), their power law spectral index,
their classification and the discovery group and date.
For the sources listed in this and subsequent tables, the original
references can be found in \cite{aharonian08}. Clearly the predominant
sources are BL Lac objects whose SEDs are peaked towards higher frequencies.
In all of these objects the TeV emission is associated with the jets.

The distinction between the objects listed in Tables 3 and 4 is not always
clear cut. Classical shell supernovae remnants are relatively easy to model
and to understand.  
Objects like those found in the vicinity of W28 are obviously
related and their subdivision into distinct sources is somewhat arbitrary.
There is particular interest in objects in which the supernovae
seems to be interacting with nearby molecular clouds; these are the best
candidates for hadronic acceleration and interaction.
Pulsar Wind Nebulae are more difficult to model and less likely to be
hadronic sources. The pulsar is generally off center and the structure 
is complex; the Crab Nebula is clearly an exception.

The Milagro sources (Table \ref{milagro}) are all detected at energies
in excess of 20 TeV. Only one has been also detected at lower energies
by IACT observatories. Since they are all extended and have no clear 
counterparts, they represent a population that is quite distinct from
the sources listed in the other tables.

Table \ref{oddsandends} lists a number of diverse but important identified sources.
Among these are the four binaries (whose parameters are also
listed in Table \ref{binaries}) \cite{parades08}. 
They do not fit into any standard class 
but these objects probably represent only a small sample of the TeV binary emitters.

In many ways the unidentified sources listed in Table \ref{dark} are most
interesting since they might represent an entirely new class of object. They
are located very close to the Galactic Plane; they seem to be
concentrated towards the Galactic Center; they are all extended; all have 
very flat spectra. They may turn out to be Pulsar Wind Nebulae that for some
reason are obscured at other wavelengths; this is the least exciting 
possibility. Other suggestions are that they might be unusual supernovae remnants,
pulsar wind nebulae, giant molecular clouds, stellar clusters with powerful
winds or Gamma Ray Burst remnants.
 
\begin{table}
\begin{tabular}{llllrll}
\hline
\tablehead{1}{r}{b}{Catalog Name}
& \tablehead{1}{r}{b}{Common Name}
& \tablehead{1}{r}{b}{Redshift}
& \tablehead{1}{r}{b}{Flux at 1 TeV\\$10^{-12}cm^{-2}s^{-1}TeV^{-1}$}
& \tablehead{1}{r}{b}{Index}
& \tablehead{1}{r}{b}{Classification}
& \tablehead{1}{r}{b}{Discovery\\(Group/Date}\\
\hline
TeV1231+124 & M87 & 0.00436 &     1 & 2.9 & FRI & HEGRA/2003\\
TeV1104+382 & Markarian 421 & 0.031 & 12-97 & 2.4-3.1 & HBL& Whipple/1992\\
TeV1654+398 & Markarian 501 & 0.034 & 0.5-100& 1.9-2.3& HBL& Whipple/1996\\
TeV2347+517 & 1ES2344+514 & 0.044 & 1-5& 2.3-2.5 & HBL  & Whipple/1998\\
TeV1136+702 & Markarian 180 & 0.045 & 0.9 & 3.3 & HBL& MAGIC/2006\\
TeV2000+651 & 1ES1959+650 & 0.048 & 4-120&2.7-2.8& HBL&Tel.Arr./2000\\
TeV0551-323 & PKS0548-323 & 0.067 & 0.3 & 2.8 & HBL& H.E.S.S./2007\\
TeV2203+423 & BL Lacertae & 0.069 & 0.3 & 3.6 & LBL& Crimea/2001\\
TeV2009-488 & PKS2005-489 & 0.071 & 0.2 & 4 & HBL & H.E.S.S./2005\\
TeV0152+017 & RGB J0152+017 & 0.080 &   & 2.95  & HBL    &H.E.S.S./2008\\
TeV1221+283 & W Comae     & 0.102 &    & 3.81  & IBL& VERITAS/2008\\
TeV2159-302 & PKS2155-304 & 0.117 & 2-3 & 2.3-2.5 & HBL& Durham/1999\\
TeV1429+427 & H1426+428   & 0.129 & 1-2 & 2.6-3.7 & HBL& Whipple/2002\\
TeV0809+524 & 1ES0806+524 & 0.138 &     &    & HBL& VERITAS/2008\\
TeV0233+203 & 1ES0229+200 & 0.140 & 0.62 & 2.5 & HBL& H.E.S.S./2006\\
TeV2359-306 & H2356-309   & 0.165 & 0.3 & 3.1 & HBL & H.E.S.S./2006\\
TeV1221+302 & 1ES1218+304 & 0.182 & 1.3 & 3.0 & HBL & MAGIC/2006\\
TeV1103-232 & 1ES1101-232 & 0.186 & 0.4 & 2.9 & HBL & H.E.S.S./2007\\
TeV0349-115 & 1ES0347-121 & 0.188 &0.45 & 3.1 & HBL & H.E.S.S./2007\\
TeV1015+495 & 1ES1011+496 & 0.212 & 0.3 & 4.0 & HBL & MAGIC/2007\\
TeV1556+112 & PG1553+113  & 0.3-04 & 0.1-0.2 & 4.0 & HBL & H.E.S.S./2006\\
TeV0219+425 & 3C66a      & 0.444 &     &   & IBL & Crimea/1998\\
TeV1256-058 & 3C279       & 0.536 &  -  & 4.1 &  FSRQ & MAGIC/2008\\
TeV0716+714 & S50716+714  & ?    &      &   & HBL & MAGIC/2008\\

\hline
\end{tabular}
\caption{ Extragalactic Sources}
\label{extragalactic}
\end{table}



\begin{table}
\begin{tabular}{llrrrrl}
\hline
\tablehead{1}{r}{b}{Object\\Catalog}
& \tablehead{1}{r}{b}{Common Name}
& \tablehead{1}{r}{b}{l$_{II}$}
& \tablehead{1}{r}{b}{b$_{II}$}
& \tablehead{1}{r}{b}{Type}
& \tablehead{1}{r}{b}{Distance\\kpc}
& \tablehead{1}{r}{b}{Discovery\\Group/Date}\\
\hline
TeV0616+225   & IC443 & 189.03 & 2.90& Shell (PWN?)  & 1.5 & MAGIC/2007 \\
TeV0852-463   &  R0852-4622 & 266.28 & 1.24 & Shell  &  10.2 & H.E.S.S./2005\\
TeV1442-625   & RCW 86 &  &  & Shell & 1  & H.E.S.S./2007 \\
TeV1714-398   & RX J1713.7-3946 & 347.28 & 0.38  & Shell & 1  & CANGAROO/2001 \\
TeV1714-382   & CTB37B & 348.65 & -0.38 & 10.2 & SNR & H.E.S.S./2006 \\
TeV1714-385   & CTB37A & 348.39 & 0.11 & 10.3 & SNR   & H.E.S.S./2008 \\
TeV1747-282   &  G0.9+0.1 &  0.87 & 0.08 & SNR (PWN?) & 8.5  & H.E.S.S./2005\\
TeV1802-233   & W28       & 6.66  & 0.27 &  Shell     & 2  & H.E.S.S./2008\\
TeV1833-105   & G21.5-0.9 & 21.5& -0.7 & Shell (PWN?) & 4   & H.E.S.S./2008\\
TeV1846-027   &  Kes 75    & 29.9& 0.0 &  SNR (PWN?)  & 6 - 19  & H.E.S.S./2008\\
TeV2323+588   & Cassiopeia A  & 111.73 & -2.1 & SNR  & 3.4 & HEGRA/2001 \\
\hline
\end{tabular}
\caption{Supernova Remnants}
\label{snr}
\end{table}



\begin{table}
\begin{tabular}{llrrrrr}
\hline
\tablehead{1}{r}{b}{Object\\Catalog}
& \tablehead{1}{r}{b}{Common Name}
& \tablehead{1}{r}{b}{Association}
& \tablehead{1}{r}{b}{l$_{II}$}
& \tablehead{1}{r}{b}{b$_{II}$}
& \tablehead{1}{r}{b}{Distance\\kpc}
& \tablehead{1}{r}{b}{Discovery\\Group/Date}\\
\hline
TeV0535+220 & Crab Nebula & M1 & 184.56 & -5.78            & 2         & Whipple/1989\\
TeV0835-463 & Vela X      &    & 263.91 & -3.01           & 0.29      & H.E.S.S./2006\\
TeV1418-610 & Kookaburra Rabbit & G313.3+0.1?  & 313.25 & 0.15 & 5.6   & H.E.S.S./2006\\
TeV1420-607 & Kookaburra Pulsar & P1420-6048   & 313.56 & 0.27 & 5.6   & H.E.S.S./2006\\
TeV1514-592 &           &MSH 15-52             & 320.33 & -1.19& 5.21  & H.E.S.S./2005\\
TeV1641-465 &           & G338.3-0.0           & 338.32 & -0.02& 8.6   & H.E.S.S./2005\\
TeV1718-385 &           &                     &  348.83 & -0.49 &4.2   & H.E.S.S./2005\\
TeV1811-193 &           &   PSR J1809-1917    & 11.18 & -0.09 & 3.7    & H.E.S.S./2007\\
TeV1826-138 &           & PSR J1826-1334      & 17.82 & -0.74 &  3.9   & H.E.S.S./2005\\
TeV1913+102 &           & PSR J1913+1011      & 44.39 & -0.07 &        & H.E.S.S./2007\\
\hline
\end{tabular}
\caption{ Pulsar Wind Nebulae}
\label{pwn}
\end{table}


\begin{table}
\begin{tabular}{lrrrlr}
\hline
\tablehead{1}{r}{b}{Object\\Catalog}
& \tablehead{1}{r}{b}{Common Name}
& \tablehead{1}{r}{b}{l$_{II}$}
& \tablehead{1}{r}{b}{b$_{II}$}
& \tablehead{1}{r}{b}{Size\\degrees}
& \tablehead{1}{r}{b}{Discovery\\Group/Date}\\
\hline
TeV1908+060 & MGRO J1908+06    & 40.16 & -0.93 & 0.5         & Milagro/2007\\
TeVJ2019+37 & MGRO J2019+37    & 75.11 & 0.54  & 1.1$\pm0.5$ & Milagro/2007\\
TeV2031+41  & MGRO J2031+41    & 79.72 & 0.94  & 3.0$\pm0.9$ & Milagro/2007 \\
\hline
\end{tabular}
\caption{Milagro Sources}
\label{milagro}
\end{table}


\begin{table}
\begin{tabular}{llrrlrl}
\hline
\tablehead{1}{r}{b}{Object\\Catalog}
& \tablehead{1}{r}{b}{Common Name}
& \tablehead{1}{r}{b}{l$_{II}$}
& \tablehead{1}{r}{b}{b$_{II}$}
& \tablehead{1}{r}{b}{Type}
& \tablehead{1}{r}{b}{Distance\\kpc}
& \tablehead{1}{r}{b}{Discovery\\Group/Date}\\
\hline
TeV0240+612 & LSI +61 303  & 135.68  & 1.09 & Binary &   2     & MAGIC/2006\\
TeV0633+058 &              & 205.66  & -1.44   & Binary? &   1.6   & H.E.S.S./2007\\
TeV1023-575 & Westerlund 2 & 284.19  & -0.39   & Stellar cluster & 8 & H.E.S.S./2007\\
TeV1302-638 & PSR 1259-63  & 304.19 & -0.99 & Binary     & 1.5 & H.E.S.S./2005\\
TeV1746-290 & Galactic Center & 359.95 & -0.05 &  Black Hole & 8.5 & CANGAROO/2004\\
TeV1746-290 & Galactic Ridge & 359.95  & -0.05 & Diffuse Source & 8.5 & H.E.S.S./2006\\ 
TeV1759-240 & W28C     & 5.7     & -0.1           & Unknown         & 2 - 4 & H.E.S.S./2008\\        
TeV1800-240 & W28B     & 5.90 & -0.36      & Molecular Cloud & 2 - 4 & H.E.S.S./2008\\
TeV1800-240 & W28A     & 6.14 & -0.63      & Molecular Cloud & 2 - 4 & H.E.S.S./2008\\
TeV1826-149 & LS 5039  & 16.88 & -1.29     &  Binary     & 2.5 & H.E.S.S./2005\\
TeV1958+352 & Cyg X-1  &  71.3& 3.1  & XRB        & 2.2  & MAGIC/2007\\ 
\hline
\end{tabular}
\caption{Binaries and Odd and Ends}
\label{oddsandends}
\end{table}

\begin{table}
\begin{tabular}{lllll}
\hline
\tablehead{1}{r}{b}{Source/Parameter}
& \tablehead{1}{r}{b}{PSR B1259-69}
& \tablehead{1}{r}{b}{LSI+61 303}
& \tablehead{1}{r}{b}{LS5039}
& \tablehead{1}{r}{b}{Cygnus X-1}\\
\hline
Type             &  B2Ve+NS     & BOVe+NS    & O6.5+BH? & O9.7Iab+BH\\
Distance (kpc)   & 1.5          & 2.0$\pm0.2$  & 2.5       & 2.2$\pm0.2$\\
Periodicity (days) & 1,237      & 25.5       & 3.9      & 5.6       \\
VHE              & Regular ?    & Irregular  & Regular  & Transient \\
Observatory      & H.E.S.S.     & MAGIC, VERITAS& H.E.S.S. & MAGIC\\
Radio Period     & 48 ms, 3.4 yr& 26.5 d, 4 yr& Steady     & Steady\\
L (X-rays) x10$^{33}$ erg/s & 0.3 - 6 & 3 - 9 & 5 -50      & 10,000\\
L (VHE gamma)x 10$^{33}$ erg/s & 2.3 &  8     &  7.8       & 12\\
Index (VHE)      & 2.7$\pm{0.2}$ & 2.6$\pm{0.2}$  &2.06$\pm{0.05}$ &3.2$\pm0.6$\\
EGRET Source    &   -          & 3EG J0241+6103& 3EG J1824-1514 & -\\
\hline
\end{tabular}
\caption{Binary Source Parameters (\cite{parades08}}
\label{binaries}
\end{table}



\begin{table}
\begin{tabular}{lrrrrr}
\hline
\tablehead{1}{r}{b}{Object//Catalog}
& \tablehead{1}{r}{b}{l$_{II}$}
& \tablehead{1}{r}{b}{b$_{II}$}
& \tablehead{1}{r}{b}{Index}
& \tablehead{1}{r}{b}{Association\\(Possible)}
& \tablehead{1}{r}{b}{Discovery\\Group/Date}\\
\hline 
TeV1303-632  & 304.24   & -0.36   &  2.4  &             & H.E.S.S./2005\\
TeV1428-608  & 314.41   & -0.14   &  2.16 &             & H.E.S.S./2008\\
TeV1614-518  & 331.52   & -0.58   &  2.46 &             & H.E.S.S./2005\\
TeV1616-509  & 332.39   & -0.14   &     &PSR J1617-5055 & H.E.S.S./2005\\
TeV1626-490  & 334.77   &  0.05   &  2.18 &             & H.E.S.S./2008\\
TeV1632-478  & 336.38   &  0.19   &  2.12 & I16320-4751 & H.E.S.S./2006\\
TeV1635-473  & 337.11   &  0.22   &  2.4  & I16358-4726 & H.E.S.S./2006\\
TeV1703-420  & 344.30   & -0.18   &  2.1  & P1702-4128  & H.E.S.S./2006\\
TeV1708-410  & 345.68   & -0.47   &  2.46 &             & H.E.S.S./2006\\
TeV1732-347  & 353.57   & -0.62   &  2.3  &             & H.E.S.S./2008\\
TeV1745-304  & 358.71   & -0.64   &  1.82 &             & H.E.S.S./2006\\
TeV1805-217  &   8.40   & -0.03   &  2.7  & G8.7-0.1    & H.E.S.S./2005\\
TeV1809-194  &  10.92   & 0.08    &       & PSR J1809-1917 &H.E.S.S./2007\\
TeV1814-178  &  12.81   & -0.03   &  2.1  & G12.82-0.02 & H.E.S.S./2005\\
TeV1835-088  &  23.24   & -0.32   &  2.5 & W41/G23.3-0.3& H.E.S.S./2005\\
TeV1838-070  &  25.18   & -0.11   &  2.27 & G25.5+0.0   & H.E.S.S./2005\\
TeV1841-056  &  26.80   & -0.2    &  2.4  &             & H.E.S.S./2008\\
TeV1857+027  &  35.96   & -0.06   &  2.39 &PSR1856+025  & H.E.S.S./2008\\
TeV1858+021  &  35.58   & -0.58   &  2.1  &             & H.E.S.S./2008\\
TeV2032+415  &  80.25   &  1.07   &  1.9 & Cyg OB2      & HEGRA/2002\\                   
 \hline
\end{tabular}
\caption{Unidentified (Dark) Sources}
\label{dark}
\end{table}


\section*{Outlook}

\subsection*{Retrospective}

Nearly fifty years ago the real pioneers of TeV astronomy,
Chudakov and Zatsepin in the U.S.S.R. and Jelley and Porter
in the British Isles,
 made their first brave venture into
this now rich field of TeV gamma-ray astronomy; it was a
major leap of faith since it was truly terra incognita. When
these early experiments were planned, there were no known
100 Mev sources; indeed in 1960 there were no known X-ray
sources and thus the known astrophysical electromagnetic
spectrum beyond the earth effectively ended in the near
ultraviolet. Although the observation of the cosmic
radiation was a clear indication that high energy particles
must be accelerated somewhere, commonsense would have
suggested that since the cosmic particles were observed to
have a steeply falling energy spectrum, it was unlikely that
the first high energy sources would be apparent at the high gamma-ray
energies where the fluxes would be very low. As early as 1962
it was pointed out that photon-photon absorption might be a
serious limitation for TeV gamma-ray astronomy (\cite{nikishov62});
fortunately the optical cosmic photon density was
overestimated and the gamma-ray horizon was not as near as
these gloomy predictions indicated. The few models of TeV
source intensity that were proposed considered only pion 
production in hadron
collisions and were speculative at best.

In fact these early pioneers were either woefully ignorant
of the astrophysics or extremely optimistic; either way they
had to have great self-confidence to feel that they had the ability to
beat the odds!

TeV Gamma-ray Astronomy is now a mature science with cutting
edge instruments, mature observatories, dedicated and
experienced adherents, and catalogs of diverse sources. The
question might then be asked why this discipline took so
long to develop for, unlike its counterparts in the X-ray
and 100 MeV gamma-ray bands, it did not have to await the
development of space technology. By 1980 the concept for a
new generation of telescope had already been proposed 
(\cite{jelleyporter};\cite{porterhill};\cite{grindlay76};
\cite{turverweekes78})

It may be that this second generation of astronomers were
not imaginative enough or were too conservative to exploit
the possibilities of these energy bands. But the slow
progress must also be at least partially because
the ground-based Cherenkov technique was not easy to
categorize and thus it was difficult for funding agencies to
fit it into their normal modes of support for astrophysical
research. At a time when "gamma-ray astronomy" was
synonymous with space astronomy (and support therefore
assumed to come from national space agencies such as NASA), there was
no natural conduit for serious funding. It was not really
high energy physics, it was not space astronomy, it was not
traditional cosmic ray astronomy and it required the dark
high remote sites traditionally associated with optical
astronomy. Even the Smithsonian Astrophysical Observatory,
which, with internal funding, had been prepared to gamble on
such ventures in the lush days of the sixties withdrew
support in 1976 when it was apparent that the early results
were not promising enough for NASA to support as providing
useful complementary observations to the space missions.

The first real impetus to develop the IACT in the USA came
about as a direct result of the activity of neutrino
astronomers who had a high profile and 
ready constituency of support from
high energy physicists. In fact it was largely pressure from
the neutrino astronomy community to build a major underwater
telescope that led the High Energy Division of the U.S.
Department of Energy to fund the pioneer effort in IACT at
the Whipple Observatory in 1982. The thinking here was that
if it could be demonstrated that there were no sources of
TeV gamma rays detected with this more sensitive, but
relatively inexpensive, technique, then there would be
little justification for the construction of the much more
expensive neutrino telescopes. The slow development of the
IACT was thus justification for the agencies to delay the
large investment necessary for the construction of large
neutrino telescopes; high energy physics funding could then
be reserved for the perceived more interesting area of high
energy particle research at accelerators. The subsequent
success of the IACT observatories has been a major impetus
for the construction of the new generation of neutrino
telescopes even though there is still not strong evidence that the
progenitor particles in most TeV gamma-ray sources are
hadrons and therefore likely neutrino producers.

\subsection{Perspective}

The present plethora of sources at TeV energies must
mean that the prospects for fruitful research are bright
for the coming epoch. If the number of participants means
anything, then one cannot but be impressed by the fact that
hundreds
of scientists (if one is to judge by the list of authors) 
have migrated into this field; this is far in excess of the
numbers at any previous epoch. 

While much has been done, there is still much to be done.
Among the potential sources yet to be detected are:

\begin{itemize}
\item Pulsars (more than one)
\item Starburst Galaxies
\item Dark Matter
\item Auger Source Counterparts
\item Gamma Ray Bursts
\end{itemize}

Although there is more than enough work to do with the
present instrumentation it is important, and indeed
inevitable, that some effort is
devoted to the development of the technology necessary for 
a new generation of
telescopes. It is easy to envisage the extension of the IACT
by simply multiplying the number of telescopes; the
difficulty is to do so economically. An N-fold increase in
the number of telescopes (and approximately in cost) only
results in an increase in sensitivity by a factor of
N$^{0.5}$. In the absence of a major technical breakthrough,
then we are entering into an era of extremely expensive
ground-based observatories; in fact the costs will begin to
be in the same ballpark as the cost of building space telescopes.

There is not a single driving scientific justification for
this major upgrade in sensitivity but rather a desire to do
better in all the areas currently being explored \cite{buckley08}.
 In the current, rather gloomy, economic climate it will be
a brave effort to seek funding in excess of a hundred
million dollars or euros without a single critical
scientific imperative (like the nature of dark matter, the
existence of dark energy or the meaning of life!). However
no such single mission objective existed for {\it Fermi} (GLAST) either.

Two such efforts of IACT construction are now 
under consideration: the largely
European CTA which aims to scale up the existing arrays and
hopes to achieve savings by mass production, and the
somewhat more innovative US AGIS which would attempt to
develop a new approach to telescope and camera design.
Some might argue that all resources could be pooled so
that one major observatory could be built with the maximum sensitivity.
However while it makes sense to have as much cooperation 
as possible between such observatories, there is a strong 
argument to be made for at least two 
independent observatories using different technologies 
(apart from the obvious one of needing
two to fully cover both hemispheres). It is very important 
in extending a discipline into a new region of parameter
space that marginal and threshold detections can be confirmed 
and that systematic effects be identified. AGILE and {\it Fermi}
are in this happy situation.
Given the wide range of phenomena that might be observed it
is unlikely that the observation requirements can be satisfied
by a single instrument without introducing serious compromises
in its design. Also since some four decades of the electromagnetic
are available, it may be sensible to concentrate and optimize
on particular bands. The large fields of view that are optimum
for the study of extended Galactic sources are generally wasteful
for the study of point sources like AGN. The mapping of supernovae
remnants at high energies requires instruments with very large
collection areas and high angular resolution.

The ongoing observations using the IACT will be complemented
by the continued operation and extension of the ground-based
particle arrays. These have already demonstrated that they are 
capable of detecting TeV sources; long integrations are possible
because of their high duty cycle. They are particularly sensitive
to extended sources. Although the anticipated detection of transient
sources has not materialized, the completion of HAWC (this symposium)
by the Milagro group 
at a high elevation site in Mexico will open new possibilities in 
this regard.

Given the fiscal realities of 2008 
it will not be surprising if, for the next few years, the ground-based
IACT community must rely on incremental improvements to existing
observatories, rather than order of magnitude scaling up
to a new generation of observatories. It is somewhat disappointing that
despite the success of the IACT, no dramatic improvement in
detection technique has been proposed. A moratorium
imposed by the bleak economy
may be advantageous in that the results obtained with the
current observatories can be thoroughly considered, new 
technologies may be taken advantage of, and innovative 
ideas for new detection strategies fostered.   

In the coming decade gamma-ray astronomy in the GeV regime will
surely be dominated by
AGILE and {\it Fermi}; complementary observations by upgraded versions
of the current ground-based observatories at energies in excess
of 100 GeV where the IACT is most sensitive will extend the
scientific impact of these missions and form a firm basis for new facilities,
both in space and on the ground.

\begin{theacknowledgments}
My research is supported by grants from the U.S. Department
of
Energy, the U.S. National Science Foundation, and the
Smithsonian Institution. 
Participation in the Heidelberg Gamma-ray 2008 Symposium 
was possible using personal funding sources.
I am grateful to my VERITAS colleagues (M.Beilicke, P.Cogan, 
D.Horan, P.Fortin, M. Hui) for supplying figures ahead
of publication and to J.Perkins for reading the manuscript.
\end{theacknowledgments}

\end{document}